\documentstyle[12pt]{article}
\def\beq{\begin{equation}}
\def\eeq{\end{equation}}
\def\bea{\begin{eqnarray}}
\def\eea{\end{eqnarray}}
\def\bq{\begin{quote}}
\def\eq{\end{quote}}

\def\AP{{\it Ann.Phys.} }

\def\BAMS{{\it Bull.Am.Math.Soc.} }

\def\CMP{{\it Comm.Math.Phys.} }

\def\HPA{{\it Helv.Phys.Acta} }

\def\NC{{\it Nuovo Cimento} }

\def\gappeq{\mathrel{\rlap {\raise.5ex\hbox{$>$}}
{\lower.5ex\hbox{$\sim$}}}}

\def\lappeq{\mathrel{\rlap{\raise.5ex\hbox{$<$}}
{\lower.5ex\hbox{$\sim$}}}}

\def\Toprel#1\over#2{\mathrel{\mathop{#2}\limits^{#1}}}

\begin{document}
\pagestyle{empty}
\begin{flushright}
{CERN-TH/2001-330}
\end{flushright}
\vspace*{5mm}
\begin{center}
{\bf BOUND STATES IN $n$ DIMENSIONS (Especially $n = 1$ and $n = 2$)}
\\
\vspace*{0.5cm}
{\bf N.N. KHURI}
\\ \vspace{0.1cm}
 Department of Physics, The Rockefeller University \\
New York, NY 10021, U.S.A.\\
\vspace*{0.3cm}
{\bf Andr\'e MARTIN}\\ \vspace{0.1cm}
Theoretical Physics Division, CERN\\
CH - 1211 Geneva 23, Switzerland\\
and\\
Laboratoire de Physique Th\'eorique ENSLAPP\\
F - 74941 Annecy-le-Vieux, France\\
\vspace{0.3cm}
{\bf Tai Tsun WU}\\ \vspace{0.1cm}
Gordon McKay Laboratory, Harvard University\\
Cambridge, MA 02138-2901, U.S.A.\\
and\\
Theoretical Physics Division, CERN\\
CH - 1211 Geneva 23, Switzerland\\
\vspace{0.2cm}
\vspace{0.2cm}
{\it Talk given by A. Martin at the Workshop\\
``Critical Stability of Few-Body Quantum Systems", Les Houches, October 2001\\
to appear in ``Few-Body Problems".}

\vspace*{0.5cm}

{\bf ABSTRACT} \\ \end{center}

\vspace*{2mm}

We stress that in contradiction with what happens in space dimensions $n
\geq 3$, there is no strict bound on the number of bound states with the
same structure as the semi-classical estimate for large coupling constant
and give, in two dimensions, examples of weak potentials with one or
infinitely many bound states. We derive bounds for one and two dimensions
which have the ``right" coupling constant behaviour for large coupling.
\vspace*{0.3cm}

\begin{flushleft} 
CERN-TH/2001-330 \\
November 2001
\end{flushleft}

\newpage

\setcounter{page}{1}
\pagestyle{plain}

\section{Introduction}

First, I would like to apologize for the fact that I am speaking on
something which, in two respects, is outside the subject of this
workshop:\\(i) I speak of two-body bound states while we speak here of
few-bodies which means at least three;\\
(ii) the most important results I present are in one- and two-space
dimensions.

My excuse for (i) is that control of two-body bound states is useful for
more
body bound states. For instance the proof that the system
proton-electron-muon is unstable uses two-body results. For (ii) it is well
known that two-space dimensions are very often realized in condensed matter
physics.

\section{Large coupling constant behaviour}

If we take a potential $gV$, the number of negative energy bound states in
$n$ dimensions is given by the semi-classical expression for large $g$
\beq
N_{total}
\simeq c_n~g^{n/2} \int d^nx \vert V^-(x)\vert^{n/2}~,
\label{one}
\eeq
where $V^-$ designates the negative part of the potential, and $c_n$ the
semi-classical constant is given by
\beq
c_n = {2^{-n}\pi^{-n/2}\over \Gamma (1+{n\over 2})}
\label{two}
\eeq

This holds for any $n$, provided $V$ is sufficiently smooth and sufficiently
rapidly decreasing at infinity. For $n = 1$, it was established by Chadan
\cite{aaa} and for general $n$ by Martin \cite{bb} and some years later by
Tamura \cite{cc}.

\section{Strict bounds for $n \geq 3$, absence of bounds for $n = 1,2$}

For $n \geq 3$, it was established by Lieb \cite{dd}, Cwikel \cite{ee} and
Rozenblum \cite{ff}, and later by others \cite{ggg} that besides the
asymptotic estimate (\ref{one}) there is a \underline{strict} bound of the
same form
\beq
N_{total} < b_n~g^{n/2} \int d^nx \vert V^-(x)\vert^{n/2}
\label{three}
\eeq
where $b_n$ is \underline{strictly} larger than $c_n$, even for very large
$n$, as shown by Glaser, Grosse and Martin \cite{hh}.

On the contrary, in one and two dimensions, the situation is drastically
different.

For $n = 1$ and $n = 2$ an arbitrarily weak attractive potential has at
least one bound state. For $n = 1$ this is easy to demonstrate using a
gaussian wave function. For $n = 2$ it is also true but more delicate
\cite{jj}. The most elegant proof is given by Yang and De Llano \cite{kk}
using a trial function $\exp -(r+r_0)^\alpha$, $\alpha$ very close to zero.

It is in fact sufficient, both for $n = 1$ and $n = 2$ to have an
arbitrarily weak, globally attractive potential, i.e., such that
\beq
\int V(x) d^nx < 0
\label{four}
\eeq
to have a bound state.

One can go further than that, i.e. construct examples in which an
arbitrarily weak potential has infinitely many bound states. One such
example, inspired by Jean-Marc Richard, is, in two dimensions
\beq
V = -\sum^\infty_{n=1} ~~g_n \delta (\vert x-x_n\vert -1)
\label{five}
\eeq
$g_n$ decreasing, $g_1$ arbitrarily small, $x_n$
adequately chosen. One can manage to have (\ref{five}) satisfying
\beq
\int \vert V\vert d^2 x < \infty~,
\label{six}
\eeq
and even
\bea
&&\int \vert V\vert \bigg(\ln (2+|x|)\bigg)^{1-\epsilon} ~~d^2x <
\infty~,\\
\nonumber &&\epsilon > 0~~{\rm arbitrarily ~~small.}
\label{seven}
\eea
One can construct other examples where the delta function is replaced by a
finite circular  ditch.

However, this or these bound states which occur for arbitrarily weak
potentials are incredibly weakly bound. We have been able to show that the
ground state  energy, $\epsilon = -\kappa^2$, satisfies, for a potential
$gV$
\beq
\kappa < \exp  \left(- {2\pi - g\int \ln^+\left({1\over x}\right) ~V_R~d^2x
-{1\over 2} g\int \vert V^-\vert d^2x
\over g\int \vert V^-\vert d^2 x}\right)
\label{eight}
\eeq
where $\ln^+(t) = \ln(t)$ for $t > 1$, =0 for $t < 1$, and $V_R$ is the
\underline{circular decreasing rearrangement} of $\vert V^-\vert$.

If some of you do not know what is a circular decreasing rearrangement, let
me say that seeing Mont Blanc it is easy to understand. You take the ``Carte
Vallot". The Mont Blanc has level lines. You replace these level lines by
circles centered at the origin enclosing the same area as the original level
lines. In this way you manufacture the rearranged Mont Blanc. You may have
to group together disconnected level lines if necessary.

In a recent preprint, Nieto \cite{lll} has calculated that in units where
$\hbar = c = 2m = 1$, a square well of radius 1, with a strength 0.1 has a
bound state with an energy of $10^{-18}$ which, incidentally, is a
particular case of (\ref{eight}) except for the non-dominant terms.

\section{Bounds on the number of bound states in one dimension}

Here and in the next sections we use extensively the fact that the number of
negative energy bound states is equal to the number of nodes of the zero
energy wave function. In one dimension let $x_1 x_2 \ldots x_K\ldots
x_{K+1} \ldots x_N$ be these nodes and assume that $x_K < 0 < x_{K+1}$.
Since the Schr\"odinger equation
$$
-{d^2\psi\over dx^2} + V(x) \psi = 0
$$
looks like a radial reduced three-dimensional equation with angular momentum
zero, we can use known bounds and apply them to the intervals $-\infty
\rightarrow x_K$, $x_{K+1} \rightarrow \infty$. So the Bargman bound
\cite{mm}
\beq
N_\ell < {1\over 2\ell+1} \int^\infty_0 r~V^-(r) dr (q)
\label{nine}
\eeq
which, for $\ell = 0$ reduces to
$$
N_0 < \int^\infty_0 r~V^-(r) dr~,
$$
gives us
$$
\matrix{N - K - 1 &< \int^\infty_{x_K} (x - x_K) V^-(x) dx \cr \cr
&< \int^\infty_0 x V^-(x) dx \hfill}
$$
and
$$
K < \int^0_{-\infty} \vert x \vert V^-(x) dx
$$
so that
\beq
N(m = 0) < 1 + \int^{+\infty}_{-\infty} \vert x\vert V^-(x) dx
\label{ten}
\eeq
Similarly, the bound obtained by one of us (A.M.) in three dimensions for
$\ell = 0$ \cite{nn}:
\beq
N (3 dim, \ell = 0) < \left[ \int^\infty_0 r^2\vert V^-\vert dr
\int^\infty_0 \vert V^-\vert dr \right]^{1/4}
\label{eleven}
\eeq
leads to
\beq
N(m = 0) < 1 + \sqrt{2} \left[ \int^{+\infty}_{-\infty} x\vert V^-\vert dx
\int^{+\infty}_{-\infty} \vert V^-\vert dx\right]^{1/4}
\label{twelve}
\eeq
This bound has the advantage that if we replace $V$ by $gV$ it is
\underline{linear} in $g$, like the semi-classical estimate for large $g$.

\section{Bounds in two-space dimensions}

We start with the \underline{radial} case for which the reduced
Schr\"odinger equation at zero energy is
\beq
\left[ -{d^2\over dr^2} + {m^2-1/4\over r^2} + V(r)\right] u = 0
\label{thirteen}
\eeq
It is in fact more convenient in the $m = 0$ case to work directly with the
non-reduced form:
\beq
-{d^2\psi\over dr^2} - {1\over r}~~{d\psi\over dr} + V(r)\psi = 0
\label{fourteen}
\eeq

Details of our aproach will be given in a future publication. Let me just
say that we integrate the Schr\"odinger from one node to the next and find
a
lower bound for the quantities
$$
\int^{x_{K+1}}_{x_K} \ln \left({x\over x_K}\right) V^-(x) x~~dx
$$
and
$$
\int^{x_{K+1}}_{x_K} \ln \left({x_{K+1}\over x}\right) V^-(x) x~~dx~~.
$$
From this we get
\beq
N(m=0) < 1+ \int^\infty_0 r\left\vert \ln \left({R\over r}\right)\right\vert
~~\vert V^-(r)\vert~~ dr
\label{fifteen}
\eeq
where $R$ is \underline{arbitrary}. If we call
\beq
I(R) = \int^\infty_0 r\vert \ln \left({R\over r}\right)\vert ~~\vert V^-
(r)\vert
dr
\label{sixteen}
\eeq
we can minimize with respect to $R$, and get
$I_{min} = I(R_0)$ such that
\beq
\int_0^{R_0} r\vert V^-(r)\vert dr = \int^\infty_{R_0} r\vert V^-(r)\vert dr
\label{seventeen}
\eeq
It is interesting to compare $I_{min}$ with the bound on $N(m=0)$ obtained
years ago by Newton \cite{oo} and Set\^o \cite{pp}.
\beq
N(m=0) < {{1\over 2} \int r~dr~r^\prime~dr^\prime~~\vert V^-(r)\vert
(V^-(r^\prime)\vert ~~\vert \ln \left({r\over r^\prime}\right)\vert \over
\int r~dr~\vert V^-(r)\vert}
\label{eighteen}
\eeq

It is easy to see, from the mean value theorem that the right-hand side of
(\ref{eighteen}) is \underline{larger} than $1/2 I_{min}$. It is less
trivial to prove that it is less or equal to $I_{min}$. So we get
\beq
{1\over 2} I_{min} < N ({\rm Newton-Set\hat o}) \leq I_{min}
\label{nineteen}
\eeq
which means that for $m=0$ our result is not quite as good, but simpler.
However, both bounds are optimal in the sense that one can approach
arbitrarily close to saturation.

For $m \not= 0$, the easiest thing to do is notice that the equation
(\ref{thirteen}) looks like a radial reduced Schr\"odinger equation in three
dimensions with an angular momentum $\ell = m - 1/2$. Thus, for $m \not= 0$
we can use the Bargmann bound, which is valid not only for integer $\ell$
but any real $\ell > -1/2$. However, if we want to calculate the TOTAL
number of bound states, we find, summing up the contributions from the
various values of $m$ that if we take a potential $gV$ we find a bound which
behaves like $g \ln g$ for large $g$, i.e., increases qualitatively faster
than the semi-classical bound. To avoid this we use a trick which was
invented many years ago by Glaser, Grosse and one of us (A.M.) \cite{hh}. We
transform Eq. (\ref{thirteen}) by making the change of variables
\beq
z = \ln r, v = r^2 V~,
\label{twenty}
\eeq
into
\beq
\left(-{d^2\over dz^2} + v(z)\right) \phi(z) = -\left(m^2-{1\over
4}\right)
\phi(z)
\label{twentyone}
\eeq
this is a one-dimensional Schr\"odinger equation whose eigenvalues are
\beq
e_i = -\left(m^2_i - {1\over 4}\right)
\label{twentytwo}
\eeq
Each $m_i$ (seen from the point of view of Eq. (\ref{thirteen})), is on a
``Regge trajectory", which means that if we disregard the $m = 0$ bound
states we have bound states at
$$
1,2,\ldots , [m_i] -1, [m_i]
$$
where $[m_i]$ is the integer part of $m_i$.

Each bound state has a multiplicity two corresponding to $\pm \vert m\vert$.
Hence the total number of $m\not= 0$ bound states is less than
\beq
2\sum m_i < 2 \times {2\over\sqrt{3}} \sum \vert e_i\vert^{1/2}~,
\label{twentythree}
\eeq
since $e_i/m^2_i \geq 3/4$. Finding a bound on the right-hand side is a
standard moment problem in one dimension first considered by Lieb and
Thirring \cite{qq}. Weidl \cite{rr} has been able to prove the inequality
$$
\sum \vert e_i\vert^{1/2}  < C \int \vert v(z)\vert dz~,
$$
and Hundertmark, Lieb and Thomas \cite{ss} have found the best possible
constant in
this inequality
\beq
\sum \vert e_i\vert^{1/2} < {1\over 2} \int^{+\infty}_{-\infty} \vert
v(z)\vert~ dz
\label{twentyfour}
\eeq

Hence using (\ref{twenty}), (\ref{twentythree}) and for instance our bound
(\ref{fifteen}) for $m = 0$  we get for the total number of bound states in
$n$ potential $gV$
\bea
N_{TOT} &<& 1 + g\int r~dr \vert\ln \left({R\over r}\right) \vert~~\vert
V^-{r}\vert \nonumber \\
&+& g {2\over\sqrt{3}} \int r~dr \vert V^-(r)\vert
\label{twentyfive}
\eea
i.e., a bound \underline{linear} in $g$.

We can get now a bound on the total number of bound states for a potential
\underline{without} symmetry: choose an origin and define
\bea
B(r) &=& - {inf\atop\theta}~~V(r,\theta) \nonumber \\ && \nonumber \\
B(r) &=& 0, ~{\rm if}~~ {inf\atop\theta}~~ V(r, \theta) > 0
\label{twentysix}
\eea
(with obvious notations in $V$).

From the monotonicity of the eigenvalues with respect to the potential we
get
\bea
N_{TOT} &<& 1 + g \int r~dr \vert \ln \left({R\over r}\right) \vert
B(r)\nonumber \\
&+& g {2\over \sqrt{3}} \int r~dr ~B(r)
\label{twentyseven}
\eea

It is clear that this bound will be good if $V$ has \underline{only one}
singular point which can be chosen as origin. Otherwise it may be very bad
or even meaningless. Our conjecture is that there must be a bound like
\bea
N_{TOT} &<& 1 + C_1 \int d^2x~~V_R(x) \left(\ln\vert{x_0\over
x}\vert\right)^+ \nonumber \\
&+& C_2 \int d^2x \vert V^-(x)\vert
\left(\ln\vert{x\over x_0}\vert\right)^+\nonumber \\
&+& C_3\int d^2x \vert V^-(x)\vert \nonumber
\eea
where $ V_R$ is the rearrangement of $V^-$.

\section{Acknowledgements}

We are grateful to K. Chadan, J.-M. Richard and W. Thirring for crucial
informations.

\end{document}